\RequirePackage{lineno}
\documentclass[aps,prl,twocolumn,showpacs,superscriptaddress]{revtex4}

\usepackage{amssymb}
\usepackage{amsmath}
\usepackage{graphicx}% Include figure files
\usepackage{dcolumn}% Align table columns on decimal point
\usepackage{epsfig}
\newcommand{\mpt}       {\mbox{$\not\!\!p_T$}}
\newcommand{\ttbar}     {\mbox{$t\overline t$}}

\newcommand{\Wbb}       {\mbox{$Wb\overline b$}}
\newcommand{\Wcc}       {\mbox{$Wc\overline c$}}

\newcommand{\Zbb}       {\mbox{$Zb\overline b$}}
\newcommand{\Zcc}       {\mbox{$Zc\overline c$}}

\begin{document}
\hspace{5.2in} \mbox{FERMILAB-PUB-11/191-E}
\title{
Search for a fourth generation $\boldsymbol{t'}$ quark in $\boldsymbol{p\bar{p}}$ collisions at $\boldsymbol{\sqrt{s}=1.96}$~TeV}
\vspace*{2.cm}

% once the note is approved for conferences, the following author will be used
% \author{The D\O\ Collaboration}
% \affiliation{URL http://www-d0.fnal.gov}

% use the official authorlist for publication
\affiliation{Universidad de Buenos Aires, Buenos Aires, Argentina}
\affiliation{LAFEX, Centro Brasileiro de Pesquisas F{\'\i}sicas, Rio de Janeiro, Brazil}
\affiliation{Universidade do Estado do Rio de Janeiro, Rio de Janeiro, Brazil}
\affiliation{Universidade Federal do ABC, Santo Andr\'e, Brazil}
\affiliation{Instituto de F\'{\i}sica Te\'orica, Universidade Estadual Paulista, S\~ao Paulo, Brazil}
\affiliation{Simon Fraser University, Vancouver, British Columbia, and York University, Toronto, Ontario, Canada}
\affiliation{University of Science and Technology of China, Hefei, People's Republic of China}
\affiliation{Universidad de los Andes, Bogot\'{a}, Colombia}
\affiliation{Charles University, Faculty of Mathematics and Physics, Center for Particle Physics, Prague, Czech Republic}
\affiliation{Czech Technical University in Prague, Prague, Czech Republic}
\affiliation{Center for Particle Physics, Institute of Physics, Academy of Sciences of the Czech Republic, Prague, Czech Republic}
\affiliation{Universidad San Francisco de Quito, Quito, Ecuador}
\affiliation{LPC, Universit\'e Blaise Pascal, CNRS/IN2P3, Clermont, France}
\affiliation{LPSC, Universit\'e Joseph Fourier Grenoble 1, CNRS/IN2P3, Institut National Polytechnique de Grenoble, Grenoble, France}
\affiliation{CPPM, Aix-Marseille Universit\'e, CNRS/IN2P3, Marseille, France}
\affiliation{LAL, Universit\'e Paris-Sud, CNRS/IN2P3, Orsay, France}
\affiliation{LPNHE, Universit\'es Paris VI and VII, CNRS/IN2P3, Paris, France}
\affiliation{CEA, Irfu, SPP, Saclay, France}
\affiliation{IPHC, Universit\'e de Strasbourg, CNRS/IN2P3, Strasbourg, France}
\affiliation{IPNL, Universit\'e Lyon 1, CNRS/IN2P3, Villeurbanne, France and Universit\'e de Lyon, Lyon, France}
\affiliation{III. Physikalisches Institut A, RWTH Aachen University, Aachen, Germany}
\affiliation{Physikalisches Institut, Universit{\"a}t Freiburg, Freiburg, Germany}
\affiliation{II. Physikalisches Institut, Georg-August-Universit{\"a}t G\"ottingen, G\"ottingen, Germany}
\affiliation{Institut f{\"u}r Physik, Universit{\"a}t Mainz, Mainz, Germany}
\affiliation{Ludwig-Maximilians-Universit{\"a}t M{\"u}nchen, M{\"u}nchen, Germany}
\affiliation{Fachbereich Physik, Bergische Universit{\"a}t Wuppertal, Wuppertal, Germany}
\affiliation{Panjab University, Chandigarh, India}
\affiliation{Delhi University, Delhi, India}
\affiliation{Tata Institute of Fundamental Research, Mumbai, India}
\affiliation{University College Dublin, Dublin, Ireland}
\affiliation{Korea Detector Laboratory, Korea University, Seoul, Korea}
\affiliation{CINVESTAV, Mexico City, Mexico}
\affiliation{FOM-Institute NIKHEF and University of Amsterdam/NIKHEF, Amsterdam, The Netherlands}
\affiliation{Radboud University Nijmegen/NIKHEF, Nijmegen, The Netherlands}
\affiliation{Joint Institute for Nuclear Research, Dubna, Russia}
\affiliation{Institute for Theoretical and Experimental Physics, Moscow, Russia}
\affiliation{Moscow State University, Moscow, Russia}
\affiliation{Institute for High Energy Physics, Protvino, Russia}
\affiliation{Petersburg Nuclear Physics Institute, St. Petersburg, Russia}
\affiliation{Instituci\'{o} Catalana de Recerca i Estudis Avan\c{c}ats (ICREA) and Institut de F\'{i}sica d'Altes Energies (IFAE), Barcelona, Spain}
\affiliation{Stockholm University, Stockholm and Uppsala University, Uppsala, Sweden}
\affiliation{Lancaster University, Lancaster LA1 4YB, United Kingdom}
\affiliation{Imperial College London, London SW7 2AZ, United Kingdom}
\affiliation{The University of Manchester, Manchester M13 9PL, United Kingdom}
\affiliation{University of Arizona, Tucson, Arizona 85721, USA}
\affiliation{University of California Riverside, Riverside, California 92521, USA}
\affiliation{Florida State University, Tallahassee, Florida 32306, USA}
\affiliation{Fermi National Accelerator Laboratory, Batavia, Illinois 60510, USA}
\affiliation{University of Illinois at Chicago, Chicago, Illinois 60607, USA}
\affiliation{Northern Illinois University, DeKalb, Illinois 60115, USA}
\affiliation{Northwestern University, Evanston, Illinois 60208, USA}
\affiliation{Indiana University, Bloomington, Indiana 47405, USA}
\affiliation{Purdue University Calumet, Hammond, Indiana 46323, USA}
\affiliation{University of Notre Dame, Notre Dame, Indiana 46556, USA}
\affiliation{Iowa State University, Ames, Iowa 50011, USA}
\affiliation{University of Kansas, Lawrence, Kansas 66045, USA}
\affiliation{Kansas State University, Manhattan, Kansas 66506, USA}
\affiliation{Louisiana Tech University, Ruston, Louisiana 71272, USA}
\affiliation{Boston University, Boston, Massachusetts 02215, USA}
\affiliation{Northeastern University, Boston, Massachusetts 02115, USA}
\affiliation{University of Michigan, Ann Arbor, Michigan 48109, USA}
\affiliation{Michigan State University, East Lansing, Michigan 48824, USA}
\affiliation{University of Mississippi, University, Mississippi 38677, USA}
\affiliation{University of Nebraska, Lincoln, Nebraska 68588, USA}
\affiliation{Rutgers University, Piscataway, New Jersey 08855, USA}
\affiliation{Princeton University, Princeton, New Jersey 08544, USA}
\affiliation{State University of New York, Buffalo, New York 14260, USA}
\affiliation{Columbia University, New York, New York 10027, USA}
\affiliation{University of Rochester, Rochester, New York 14627, USA}
\affiliation{State University of New York, Stony Brook, New York 11794, USA}
\affiliation{Brookhaven National Laboratory, Upton, New York 11973, USA}
\affiliation{Langston University, Langston, Oklahoma 73050, USA}
\affiliation{University of Oklahoma, Norman, Oklahoma 73019, USA}
\affiliation{Oklahoma State University, Stillwater, Oklahoma 74078, USA}
\affiliation{Brown University, Providence, Rhode Island 02912, USA}
\affiliation{University of Texas, Arlington, Texas 76019, USA}
\affiliation{Southern Methodist University, Dallas, Texas 75275, USA}
\affiliation{Rice University, Houston, Texas 77005, USA}
\affiliation{University of Virginia, Charlottesville, Virginia 22901, USA}
\affiliation{University of Washington, Seattle, Washington 98195, USA}
\author{V.M.~Abazov} \affiliation{Joint Institute for Nuclear Research, Dubna, Russia}
\author{B.~Abbott} \affiliation{University of Oklahoma, Norman, Oklahoma 73019, USA}
\author{B.S.~Acharya} \affiliation{Tata Institute of Fundamental Research, Mumbai, India}
\author{M.~Adams} \affiliation{University of Illinois at Chicago, Chicago, Illinois 60607, USA}
\author{T.~Adams} \affiliation{Florida State University, Tallahassee, Florida 32306, USA}
\author{G.D.~Alexeev} \affiliation{Joint Institute for Nuclear Research, Dubna, Russia}
\author{G.~Alkhazov} \affiliation{Petersburg Nuclear Physics Institute, St. Petersburg, Russia}
\author{A.~Alton$^{a}$} \affiliation{University of Michigan, Ann Arbor, Michigan 48109, USA}
\author{G.~Alverson} \affiliation{Northeastern University, Boston, Massachusetts 02115, USA}
\author{G.A.~Alves} \affiliation{LAFEX, Centro Brasileiro de Pesquisas F{\'\i}sicas, Rio de Janeiro, Brazil}
% ...
%
% visitor_addresses.tex                        6 April 2011
%  available symbols are:
%  $\ast, \dag, \ddag, \S, \P, $\|$, $\ast\ast$, \dag\dag, \ddag\ddag ,\#
%
\collaboration{The D0 Collaboration\footnote{with visitors from
%{alton}
$^{a}$Augustana College, Sioux Falls, SD, USA,
%{burdin}
$^{b}$The University of Liverpool, Liverpool, UK,
%{haas,partridge}
$^{c}$SLAC, Menlo Park, CA, USA,
%{hesketh}
$^{d}$University College London, London, UK,
%{luna-garcia}
$^{e}$Centro de Investigacion en Computacion - IPN, Mexico City, Mexico,
%{podesta-lerma}
$^{f}$ECFM, Universidad Autonoma de Sinaloa, Culiac\'an, Mexico,
and 
%{weber}
$^{g}$Universit{\"a}t Bern, Bern, Switzerland.
%{garcia-guerra}
%$^{?}$UPIITA-IPN, Mexico City, Mexico,
%{hooper}
%$^{?}$Visitor from Bradley University, Peoria, IL, USA.
%{kozminski}
%$^{?}$}Visitor from Lewis University, Romeoville, IL, USA.
%{deceased}
%$^{\ddag}$Deceased.
}} \noaffiliation
\vskip 0.25cm

\date{April 22, 2011}
           
%--------------------------------------------------------

\begin{abstract}
We present a search for pair production of a fourth generation $t'$ quark and its antiparticle, followed by their decays to a $W$ boson and a jet, based on an integrated luminosity of 5.3~fb$^{-1}$ of proton-antiproton collisions at $\sqrt{s}=1.96$~TeV collected by the D0 Collaboration at the Fermilab Tevatron Collider. We set upper limits on the $t'\bar{t'}$ production cross section that exclude at the 95\%~C.L. a $t'$ quark that decays exclusively to $W$+jet with a mass below 285~GeV.  We observe a small excess in the $\mu$+jets channel which reduces the mass range excluded compared to the expected limit of 320 GeV in the absence of a signal.
\end{abstract}


\maketitle

\newpage
%\begin{pagewiselinenumbers}

%\section{Introduction}

Measurements of the partial width of the $Z$ boson to invisible final states at LEP exclude the existence of a fourth neutrino flavor with a mass less than half the $Z$ boson mass~\cite{LEPZ}. However, this does not exclude the existence of a fourth generation of fermions as long as its neutrino is more massive. Precision electroweak data favor a small mass splitting between the up-type quark of this fourth generation, $t'$, and its down-type partner, $b'$, so that $m(t')-m(b')<m(W)$~\cite{Plehn}. Provided there is moderate mixing between the new fourth generation and the first three generations, the $t'$ quark will predominantly decay to $Wq$, where $q$ includes all standard model down-type quarks. 

We report on a search for a fourth generation $t'$ quark that is produced in proton-antiproton collisions together with its antiparticle. We assume that the $t'$ quark is a narrow state that always decays to $Wq$. This search is also sensitive to other new particles that are pair produced and decay to a $W$ boson plus a jet. We select lepton+jets final states with one isolated electron or muon with high transverse momentum ($p_T$), a large imbalance in transverse momentum ($\mpt$), and at least four jets corresponding to events in which one of the $W$ bosons decays to leptons and the other $W$ boson decays to quarks. A similar search has been carried out by the CDF Collaboration in 0.8 fb$^{-1}$ of integrated luminosity~\cite{CDF}. 

%\section{The D0 Detector}

The D0 detector consists of central tracking, calorimeter, and muon systems~\cite{run1det,run2det}. The central tracking system is located inside a 2~T superconducting solenoidal magnet. Central and forward preshower detectors are located just outside of the coil and in front of the calorimeters. The liquid-argon/uranium calorimeter is divided into a central section covering pseudorapidity $|\eta|<1.1$ and two end calorimeters extending $\eta$ coverage to $4.2$. The calorimeter is segmented longitudinally into electromagnetic, fine hadronic, and coarse hadronic sections with increasingly coarser sampling. The muon system, located outside the calorimeter, consists of one layer of tracking detectors and scintillation trigger counters inside 1.8~T toroidal magnets and two similar layers outside the toroids. A three-level trigger system selects events that are recorded for offline analysis.

%\section{Data Samples and Monte Carlo Simulation}

%\section{Event Selection}
This analysis is based on data corresponding to an integrated luminosity of 5.3~fb$^{-1}$, collected by the D0 Collaboration at the Fermilab Tevatron proton-antiproton collider at a center of mass energy of $\sqrt{s}=1.96$~TeV. 
Events must satisfy one of several trigger conditions, all requiring an electron or muon with high transverse momentum, in some cases in conjunction with one or more jets. For all events, the $p\overline{p}$ collision point must be reconstructed with at least three tracks and located within 60 cm of the center of the detector along the beam direction. 
Jets are reconstructed using a midpoint cone algorithm~\cite{run2jets} with cone size $\Delta R=\sqrt{(\Delta\eta)^2+(\Delta\phi)^2}=0.5$, where $\phi$ is the azimuth, and must have at least two reconstructed tracks within the jet cone. The jet energy is corrected on average to the total energy of all particles emitted inside the jet cone. Jets in simulated events are adjusted to reproduce the reconstruction efficiency and energy resolution and response observed in data. All events must have at least four jets with $|\eta|<2.5$, $p_T>40$~GeV for the leading jet, and $p_T>20$~GeV for all other jets.
The momentum carried away by neutrinos is inferred from the $\mpt$, computed from the energies in the cells of the electromagnetic and fine hadronic calorimeters and adjusted for the energy corrections applied to the reconstructed jets and electrons and for the momentum of any reconstructed muons, taking into account their energy loss in the calorimeter. 

Electrons are identified as clusters of energy depositions in the calorimeter that are isolated from other energy deposits. The electromagnetic section of the calorimeter must contain 90\% of their energy, and the energy deposition pattern must be consistent with that of an electromagnetic shower. Every electron must be matched to a reconstructed track with $p_T>5$~GeV. For the $e$+jets channel, we require exactly one electron with $p_T>20$~GeV and $|\eta|<1.1$ that originates from the $p\overline{p}$ collision point. We also require $\mpt>20$~GeV and $|\Delta\phi(e,\mpt)| > 2.2 - 0.045\cdot\mpt/\mbox{GeV}$, where $\Delta\phi(e,\mpt)$ is the azimuthual angle between electron and $\mpt$, to reject events with jets that are misidentified as electrons. 

Muons are defined as tracks reconstructed in the muon system matched to tracks in the central tracker. Muons must be separated from jets and isolated in the calorimeter and in the tracker. For the $\mu$+jets channel, we require exactly one muon with $p_T>20$~GeV and $|\eta|<2$ that originates from the $p\overline{p}$ collision point. The invariant mass of the selected muon and any other muon must be less than 70~GeV or more than 110~GeV to reject $Z(\to\mu\mu)$+jets events. We require \mpt$> 25$~GeV and $|\Delta\phi(\mu,\mpt)| > 2.1 - 0.035 \cdot \mpt/\mbox{GeV}$ to reject events with mismeasured muons.
More details about the lepton+jets event selection can be found in Ref.~\cite{ljxsec}.

%\section{Background Model}\label{sec:bkg}

The two main standard model processes that produce events with an isolated lepton, $\mpt$, and at least four jets are $t\overline{t}$ and $W$+jets production. The third most important source of events arises from mismeasured multijet events in which a jet is misidentified as an electron or a muon from heavy flavor decay appears isolated. Single top quark, $Z$+jets, and diboson production can also give rise to such final states but have much smaller cross sections and/or acceptances. 

We use {\sc alpgen}~\cite{alpgen} to simulate $t\overline t$ production with the top quark mass set to 172.5~GeV and generate additional jets from parton showers with {\sc pythia}~\cite{pythia}. We normalize the $t\overline t$ sample to the theoretical $t\overline t$ production cross section of $7.48^{+0.56}_{-0.72}$~pb~\cite{topxsec}. 
Samples of $W$+jets events are generated using {\sc alpgen} and {\sc pythia} with a jet-matching algorithm, following the MLM prescription~\cite{jetmatching}. Three subsamples are generated: \Wbb, \Wcc, and $W$+light partons. The $Wc$ subprocesses are included in the $W$+light parton sample with massless charm quarks. We fix the relative normalization of \Wbb, \Wcc, and $W$+light parton events to match NLO cross sections~\cite{VVxsec}. 
The $Z$($\rightarrow ee$, $\mu\mu$, $\tau\tau$)+jets samples are generated with {\sc alpgen} and {\sc pythia} and broken up into \Zbb, \Zcc, and $Z$+light parton samples in the same way as the $W$+jets samples. We fix their relative normalization to NLO predictions and normalize the total $Z$ boson sample to the NNLO cross section~\cite{Zxsec}. 
We simulate single top quark production using the {\sc comphep}-{\sc singletop}~\cite{comphep} Monte Carlo event generator with the top quark mass set to 172.5~GeV and normalize to the NNLO cross section with NNNLO threshold corrections in the $s$ and $t$-channels of 3.3~pb~\cite{txsec}. Diboson samples are generated with {\sc pythia}. Their NLO cross sections are 12.3~pb for $WW$, 3.7~pb for $WZ$, and 1.4~pb for $ZZ$ production~\cite{VVxsec}. The CTEQ6L1 parton distribution functions~\cite{cteq} are used for all Monte Carlo samples. We simulate detector effects using the {\sc geant}~\cite{geant} program. Events from random collisions are added to all simulated events to account for detector noise and additional $p\bar{p}$ interactions. The events are reconstructed with the same program as the data.

To define the background model, we estimate the number of multijet events that enter the final data sample using a data driven method~\cite{MatrixMethod}. We compute the number of multijet events in the $e$+jets and $\mu$+jets samples separately. We then subtract the multijet and all other backgrounds, except the $W$+jets background, from the data, based on their calculated cross sections, and normalize the $W$+jets contribution to the remaining number of events. This corresponds to scaling the total number of $W$+jets events expected by a factor $1.3$, which is consistent with NLO expectations. Table~\ref{tab:sample} summarizes the resulting composition of the data sample. When we test for the presence of a $t'$ quark signal, we fix the relative normalizations of the $W$+jets, $Z$+jets, single top quark, and diboson backgrounds, as given in Table~\ref{tab:sample}, but float their overall normalization. 

\begin{table}[ht]
\caption{Composition of the final data sample with systematic uncertainties. The number of $W$+jets events is chosen to equalize the total number of events observed and expected.}\
\centering
\begin{ruledtabular}
\begin{tabular}{lcc}
Source & $e$+jets & $\mu$+jets \\
\hline
\ttbar\ production      & ~678$\pm$76 & ~508$\pm$55 \\
Single $t$ production   & ~~12$\pm$4 & ~~~8$\pm$3 \\
$W$+jets                & ~503$\pm$87 & ~648$\pm$59 \\
$Z$+jets                & ~~41$\pm$7 & ~~40$\pm$7 \\
$WW$, $WZ$, $ZZ$+jets   & ~~25$\pm$5 & ~~21$\pm$5 \\
Multijets               & ~173$\pm$42 & ~~43$\pm$18 \\
\hline
Data                    & 1431 & 1268 \\
\end{tabular}
\label{tab:sample}
\end{ruledtabular}
\end{table}

%\section{Analysis}

To simulate the signal, we use $t'\bar{t'}$ production in {\sc pythia} and force the decay $t'\rightarrow Wb$. However, since we do not identify $b$ jets in this analysis, our results are also applicable to $t'$ quarks decaying to a $W$ boson and a light down-type quark. We generate events at 13 $t'$-mass values between 200 and 500~GeV. We set the total width of the $t'$ quark to 10~GeV. This is smaller than the resolution for reconstructing the $t'$ mass, which ranges between 50~GeV at $m_{t'}=200$~GeV and 100~GeV at $m_{t'}=500$~GeV. Therefore, the exact value of the width does not affect the analysis. 

\begin{figure}[htp]
\epsfxsize=3.5in\epsfbox{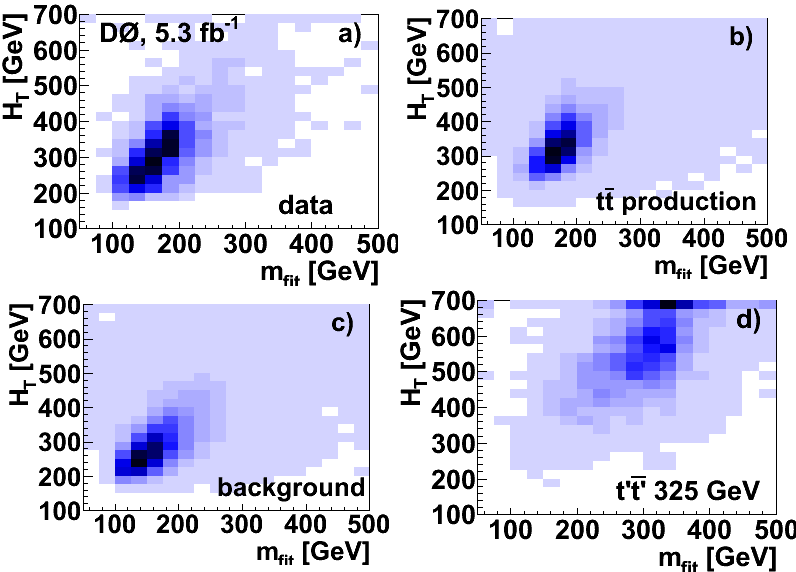}
\caption{$H_T$ versus $m_\mathrm{fit}$ for (a) data, (b) $t\overline t$-production, (c) other background, and (d) $t'\bar{t'}$ signal with $m(t')=325$~GeV. The bins at the upper and right edges of the plots also contain overflows.
\label{fig:2dhis}}
\end{figure}

We define $H_T$ as the scalar sum of $\mpt$ and of the transverse momenta of all jets and the charged lepton. A kinematic fit to the $t'\overline{t'}\rightarrow \ell\nu b q\overline{q}'\overline{b}$ hypothesis reconstructs the mass $m_\mathrm{fit}$ of the $t'$ quark. We use the two-dimensional histograms of $H_T$ versus $m_{\mathrm{fit}}$ to test for the presence of signal in the data and to compute 95\%~C.L. upper limits on the $t'\bar{t'}$ production cross section as a function of $t'$-mass. Figure~\ref{fig:2dhis} shows the scatter plots observed in data and expected from $t'\bar{t}'$ production, $t\bar{t}$ production, and from all other background sources. For each hypothesized value of the $t'$ mass, we fit the data to background-only and to signal+background hypotheses. We then use the likelihood ratio $L = -2\,\mathrm{log}(P_{S+B}/P_B)$ as the test statistic, where $P_{S+B}$ is the Poisson likelihood to observe the data under the signal+background hypothesis and $P_B$ is the Poisson likelihood to observe the data under the background-only hypothesis. For the background-only hypothesis, we fit three components to the data: $t\bar{t}$ production constrained to its theoretical cross section, the multijets background constrained to the number of events given in Table~\ref{tab:sample} and $W$+jets and all other backgrounds in the proportions given in Table~\ref{tab:sample}. For the signal+background fit we add the $t'\bar{t'}$ cross section as a parameter to the fit. The fit can discriminate between background and signal contributions because their distributions in the $H_T$ and $m_\mathrm{fit}$ variables are different. For each hypothesis we also vary the systematic uncertainties given in Table~\ref{tab:syst} subject to a Gaussian constraint to their prior values to maximize the likelihood ratio~\cite{Collie}.

We use the $CL_s$ method~\cite{CLs} to determine the cross section limits. Using pseudoexperiments, we determine the probability to measure values of $L$ that are larger than the value observed in the data sample for a $t'$ signal, $CL_{s+b}$, and for no $t'$ signal, $CL_b$. The value of the $t'$ pair production cross section for which $1-CL_{s+b}/CL_b=0.95$ is the 95\%~C.L. upper limit. We repeat this procedure for each $t'$ mass point. 

Table~\ref{tab:syst} summarizes the sources of systematic uncertainties included in the limit calculation. The first four uncertainties affect the normalization of the components of our signal and background models. All other uncertainties affect the selection efficiency. When estimating the effect of uncertainties in the jet energy scale, the jet identification efficiency, and the jet energy resolution, we also vary the shapes of the $H_T$ and $m_\mathrm{fit}$ distributions. No uncertainties are given for the $W$+jets background because its normalization is a free parameter of the fit. 

\begin{table}[ht]
\caption{\label{tab:syst} Summary of systematic uncertainties above 1\%. Some values vary with channel and with the time at which the data were taken. The numbers give the range for the size of the uncertainties.}
\begin{ruledtabular}
\begin{tabular}{lccc}
Source                   & $t'\bar{t}'$ & $t\bar{t}$ & multijets \\
\hline
$t\bar{t}$ cross section & ---          & 9\%        & ---\\
Multijets normalization  & ---          & ---        & (25--50)\% \\
Integrated luminosity    & 6.1\%        & 6.1\%      & ---\\
MC model                 & ---          & 4.3\%      & ---\\
Trigger efficiency       & $\leq$5\%    & $\leq$5\%  & ---\\
$p\bar{p}$ collision point reconstruction & 1.6\%        & 1.6\%      & ---\\
Lepton identification    & (3--4)\%     & (3--4)\%      & ---\\
Jet energy calibration   & (1--2)\%     & (2--5)\%      & ---\\
Jet energy resolution    & (1--2)\%     & (2--3)\%      & ---\\
Jet identification       & 1\%          & (1--3)\%      & ---\\
\end{tabular} 
\end{ruledtabular}
\end{table}

\begin{figure}\tt
\epsfxsize=3.5in\epsfbox{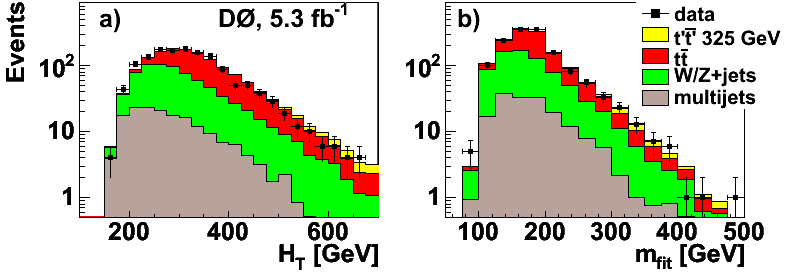}
\epsfxsize=3.5in\epsfbox{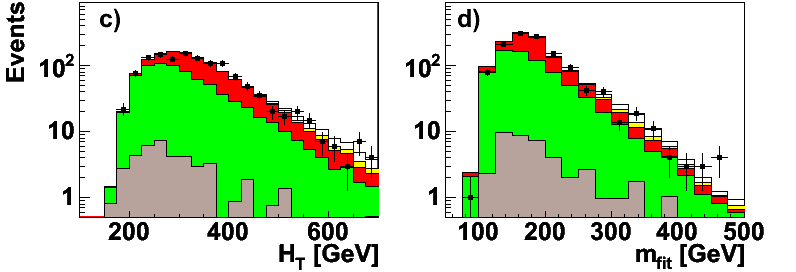}
\caption{Distributions of (a) $H_T$ and (b) $m_\mathrm{fit}$ for $e$+jets data 
and (c) $H_T$ and (d) $m_\mathrm{fit}$ for $\mu$+jets data compared with expectations. The $W/Z$+jets category also includes single top quark and diboson production. The $t'\bar{t'}$ signal is normalized to the expected yield. The unfilled histograms in (c) and (d) show the distributions with the best fit $t'\bar{t'}$-production cross section.
\label{fig:1D}}
\end{figure}

We first analyze the $e$+jets and $\mu$+jets data separately. Figure~\ref{fig:1D} shows the distributions of $H_T$ and $m_\mathrm{fit}$ from the standard model backgrounds and a 325~GeV $t'$ quark signal compared to data. There is no visible excess in the $e$+jets data. In the $\mu$+jets data we observe a small excess of events over standard model expectations. We can fit the data best with a $t'\bar{t'}$ production cross section of $3.2\pm1.1$ times the theoretical cross section for a $t'$ quark mass of 325~GeV. The value of $1-CL_b$ for the data gives the probability of getting a local deviation of at least this size from the standard model expectation in the absence of physics beyond the standard model. We find a $p$ value of 0.007, corresponding to 2.5 Gaussian-equivalent standard deviations. 

\begin{figure}[ht]
\epsfxsize=1.6in\epsfbox{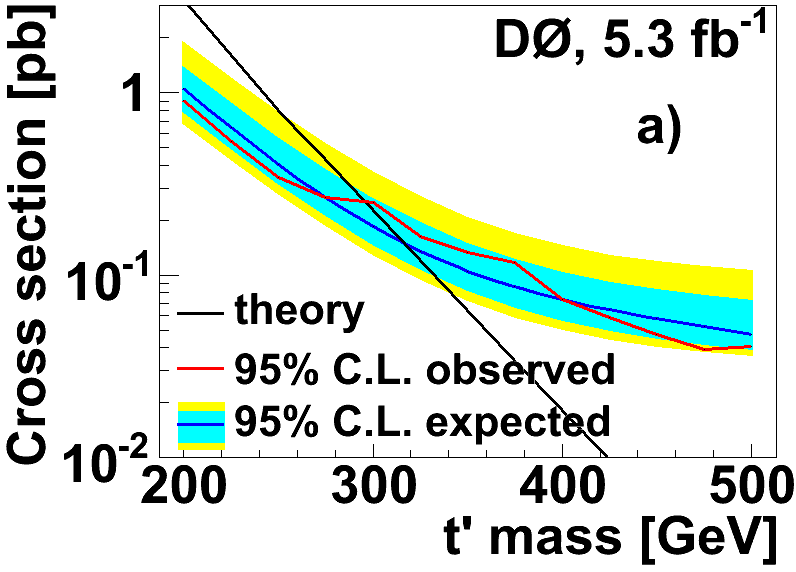}
\epsfxsize=1.6in\epsfbox{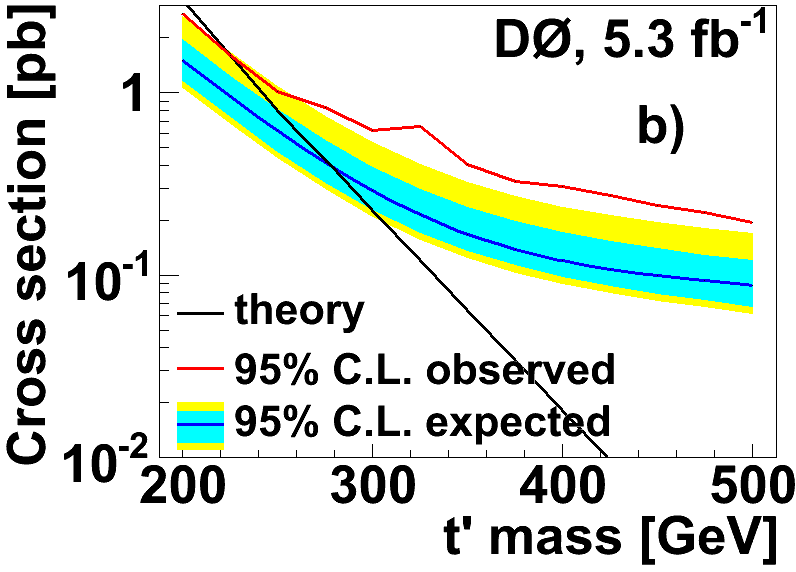}
\caption{Observed and expected upper limits and predicted values for the $t'\bar{t'}$ production cross section as a function of the mass of the $t'$ quark for (a) $e$+jets, (b) $\mu$+jets. The shaded regions around the expected limit represent the $\pm$1 and $\pm$2 standard deviation bands.\label{fig:limits}}
\end{figure}

\begin{figure}[ht]
\epsfxsize=3in\epsfbox{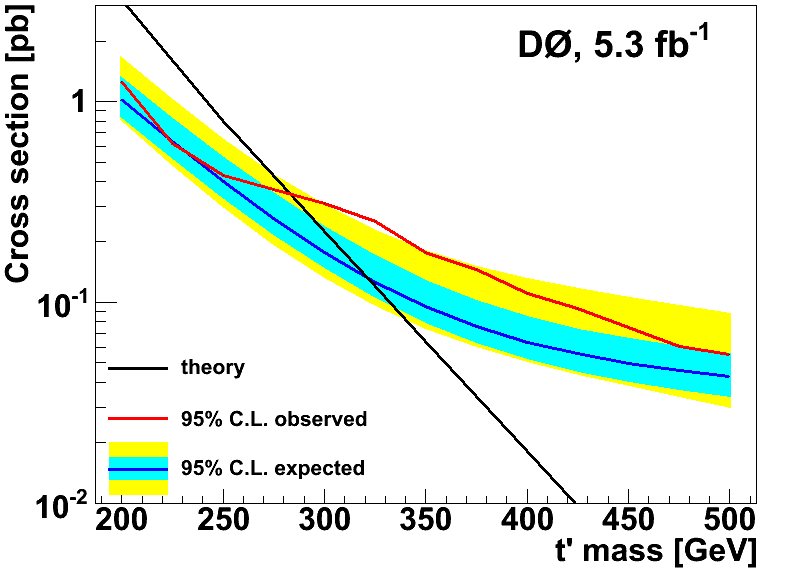}
\caption{Same as Fig.~\ref{fig:limits} but for both channels combined.
\label{fig:combined_limits}}
\end{figure}

Figure~\ref{fig:limits} shows the resulting cross section limits compared to the limits expected in the absence of $t'\bar{t'}$ production and to the predicted NLO $t'$ pair production cross section~\cite{tprime_xsec} as a function of the $t'$ mass. We expect to be able to exclude $t'\bar{t'}$ production for $t'$ quark masses below 315~GeV in the $e$+jets channel and below 280~GeV in the $\mu$+jets channel. The observed cross section limit allows us to exclude $t'\bar{t'}$ production for $t'$ quark masses at the 95\%~C.L. below 295~GeV in the $e$+jets channel and below 225~GeV in the $\mu$+jets channel. Combining $e$+jets and $\mu$+jets data as shown in Fig.~\ref{fig:combined_limits}, we expect to exclude $t'\bar{t'}$ production for $t'$ quark mass values below 320~GeV. Based on the observed limits we can exclude at the 95\%~C.L. $t'\bar{t'}$ production for $t'$ quark masses below 285~GeV. We achieve the best fit to the data with a $t'\bar{t'}$ production cross section of $1.1\pm0.5$ times the theoretical cross section for a $t'$ quark mass of 325~GeV which gives a $p$ value of 0.015, corresponding to 2.2 standard deviations from zero. 

%\section{Conclusion}

In conclusion, we searched for pair production of a $t'$ quark and its antiparticle followed by their decays into a $W$ boson and a jet. We do not see a signal consistent with $t'\bar{t'}$ production, although we observe a small excess of events in the $\mu$+jets channel. Combining the $e$+jets and $\mu$+jets channels, we exclude at 95\%~C.L. $t'\bar{t'}$ production for $t'$ quark mass values below 285~GeV. 

%\section*{Acknowledgements}
% acknowledgement.tex                             6 April 2011
%
We thank the staffs at Fermilab and collaborating institutions,
and acknowledge support from the
DOE and NSF (USA);
CEA and CNRS/IN2P3 (France);
FASI, Rosatom and RFBR (Russia);
CNPq, FAPERJ, FAPESP and FUNDUNESP (Brazil);
DAE and DST (India);
Colciencias (Colombia);
CONACyT (Mexico);
KRF and KOSEF (Korea);
CONICET and UBACyT (Argentina);
FOM (The Netherlands);
STFC and the Royal Society (United Kingdom);
MSMT and GACR (Czech Republic);
CRC Program and NSERC (Canada);
BMBF and DFG (Germany);
SFI (Ireland);
The Swedish Research Council (Sweden);
and
CAS and CNSF (China).
%

%\end{pagewiselinenumbers}

%--------------------------------------------------------

\end{document}